\hfuzz 2pt
\font\titlefont=cmbx10 scaled\magstep1
\magnification=\magstep1

\def\asymptotic#1\over#2{\mathrel{\mathop{\kern0pt #1}\limits_{#2}}}

\null
\vskip 1.5cm
\centerline{\titlefont OPEN SYSTEM APPROACH}
\medskip
\centerline{\titlefont TO NEUTRINO OSCILLATIONS}
\vskip 2.5cm
\centerline{\bf F. Benatti}
\smallskip
\centerline{Dipartimento di Fisica Teorica, Universit\`a di Trieste}
\centerline{Strada Costiera 11, 34014 Trieste, Italy}
\centerline{and}
\centerline{Istituto Nazionale di Fisica Nucleare, Sezione di 
Trieste}
\vskip 1cm
\centerline{\bf R. Floreanini}
\smallskip
\centerline{Istituto Nazionale di Fisica Nucleare, Sezione di 
Trieste}
\centerline{Dipartimento di Fisica Teorica, Universit\`a di Trieste}
\centerline{Strada Costiera 11, 34014 Trieste, Italy}
\vskip 2cm
\centerline{\bf Abstract}
\smallskip
\midinsert
\narrower\narrower\noindent
Neutrino oscillations are studied in the general framework of open quantum
systems by means of extended dynamics that take into account possible
dissipative effects. These new phenomena induce modifications in the 
neutrino oscillation pattern that in general can be parametrized by means
of six phenomenological constants. Although very small, stringent bounds 
on these parameters are likely to be given by future planned 
neutrino experiments.
\endinsert
\bigskip
\vfil\eject

{\bf 1. INTRODUCTION}
\medskip

A large variety of open quantum systems can be modeled as being subsystems
in interaction with large environments. The time evolution of the total 
system, subsystem plus environment, is unitary and follows the standard 
rules of quantum mechanics. However, the dynamics of the subsystem alone, 
obtained by eliminating the environment degrees of freedom, is no longer 
unitary, as it develops dissipation and irreversibility.[1-3]

When there are no initial correlations between subsystem
and environment and their mutual interactions can be considered weak, 
the resulting subdynamics can be described in terms of so-called quantum 
dynamical semigroups. These are time-evolution maps that encode very general 
physical requirements,  like entropy increase, forward in time composition 
law (semigroup property) and complete positivity; these properties are 
essential for the correct physical interpretation of the subdynamics.

Although this description of open quantum systems has been originally 
developed in the framework of quantum optics,[4-6] 
it is very general and can 
be applied to model a variety of different phenomena. Recently, it has 
been adopted to study the effects of dissipation and irreversibility in 
various particle physics phenomena.[7-14] 

The original motivations for such investigations were based on quantum
gravity effects:\hfill\break[15-20] due to the quantum fluctuations 
of the gravitational 
field and the appearance of virtual black holes, spacetime becomes 
``foamy'' at Planck's scale, 
leading to possible loss of quantum coherence. From a more fundamental 
point of view, also string theory could lead 
to similar effects in the low energy domain.[21] 
Indeed, subdynamics described 
by quantum dynamical semigroups are the result of the interaction 
with a ``gas'' of D0-branes at Planck's temperature, 
obeying infinite statistics.[7]
Nevertheless, since not enough details about the ``microscopic'' dynamics 
are known to allow precise estimates of the magnitude of these new effects,
the description of dissipative phenomena that we shall discuss below 
should be thought as being phenomenological in nature. 

These new, non-standard effects are very small, since they are suppressed 
by inverse powers of the Planck mass, 
as a rough dimensional estimate suggests, 
and therefore very difficult to observe in practice. 
However, there are particular situations, involving interference
phenomena, in which they might be in the reach of present or
future experiments. 
Indeed, detailed studies of neutral meson systems,[8, 10, 13] and 
neutron interferometry,[14] using quantum dynamical semigroups have already 
been performed and order of magnitude limits on some of these 
dissipative effects have been derived 
using available experimental data.[9, 11, 14]
One of the most interesting outcome of these investigations is that 
future experiments, in particular those involving
correlated neutral mesons, should be able to ascertain with high accuracy
the presence of such dissipative phenomena.

Neutrino physics is certainly another obvious place 
where to look for non-standard
effects. Many neutrino experiments are presently taking data and other will
start operating in the near future, so that it appears timely to discuss 
in detail to what extent dissipation can affect those observations.

We shall limit our considerations to the vacuum oscillations of two species 
of neutrinos. In this case, possible dissipative effects can be parametrized
in terms of six phenomenological constants that modify the pattern of the
transition probability $\cal P$ among the two neutrino flavours, 
by introducing exponential dumping factors. Although the explicit expression
of $\cal P$ is in general rather complicated, in the generic case its 
asymptotic (large time) behaviour turns out to be independent from the 
mixing angle. Various approximated expressions for $\cal P$ will also be 
discussed; they can be of help in fitting the experimental data. 
Finally, in the last section we shall present a discussion
on a possible physical mechanism that could be at the origin of 
the dissipative phenomena.

\vskip 2cm

\line{{\indent\bf 2. QUANTUM DYNAMICAL SEMIGROUPS AND NEUTRINO}\hfill}
\line{\phantom{\bf\indent 2. }{\bf OSCILLATIONS}\hfill}

\medskip

Quite in general, states of a quantum system evolving in time can be 
described by a density matrix $\rho$; this is a hermitian, 
positive operator, {\it i.e.}
with positive eigenvalues, and constant trace. 
We shall analyze the evolution
of neutrinos created in a given flavour by the weak interactions and 
subsequently detected at a later time. Assuming the neutrinos 
to be ultrarelativistic, the
study of the transition probability for the original tagged neutrinos to be
found in a different flavour can be performed using an effective 
description;[22-24] further, for simplicity, we shall limit our 
considerations to the mixing of two neutrino species. 
Then the neutrino system can be modeled by means
of a two-dimensional Hilbert space, taking as basis states 
the two mass eigenstates.

With respect to this basis, the two flavour states, that conventionally we
shall call ``$\nu_e$'' and ``$\nu_\mu$'', are represented 
by the following
$2\times2$ matrices:
$$\eqalignno{
&\rho_{\nu_e}=\left(\matrix{\cos^2\theta & \cos\theta\,\sin\theta\cr
                           \cos\theta\,\sin\theta & \sin^2\theta}\right)\ ,
&(2.1a)\cr
&\null\cr
&\rho_{\nu_\mu}=\left(\matrix{\sin^2\theta & -\cos\theta\,\sin\theta\cr
                           -\cos\theta\,\sin\theta & \cos^2\theta}\right)
\equiv 1-\rho_{\nu_e}\ , &(2.1b)}
$$
where $\theta$ is the mixing angle. 

As explained in the introductory remarks, our analysis is based on the 
hypothesis that the evolution in time of the neutrino state $\rho$ 
is given by a quantum dynamical semigroup, {\it i.e.} by a completely 
positive, trace-preserving family of linear maps: $\rho(0)\mapsto\rho(t)$. 
These maps are generated by equations of the following
form:
$$
{\partial\rho(t)\over \partial t}= -i H_{\rm eff}\ \rho(t)+i\rho(t)\,
H_{\rm eff} + L[\rho(t)]\ .\eqno(2.2)
$$
The first two terms in the r.h.s. of this equation are the standard 
quantum mechanical ones, that give rise to the traditional description of
neutrino oscillations. They contain the effective (time-independent) 
hamiltonian $H_{\rm eff}$; neglecting effects due to possible neutrino 
instability, it can be taken to be hermitian. 
The additional piece $L[\rho]$ is a linear
map, whose form is completely fixed by the conditions of complete
positivity and trace conservation:[1]
$$
L[\rho]=-{1\over2}\sum_j\Big(A^\dagger_jA_j\,\rho +
\rho\, A^\dagger_jA_j\Big)\ 
+\sum_j A_j\,\rho\, A^\dagger_j\ , \eqno(2.3)
$$
where the operators $A_j$ must be such that $\sum_j A^\dagger_j A_j$ is a
well-defined $2\times 2$ matrix. 
The additional requirement of entropy increase
can be easily implemented by taking the $A_j$ to be hermitian.[8]
It should be stressed that in absence of $L[\rho]$, pure states 
({\it i.e.} states of the form $|\psi\rangle\langle\psi|$) would be 
transformed into pure states. Only when the extra piece $L[\rho]$ is also
present, $\rho(t)$ becomes less ordered in time due to a mixing-enhancing
mechanism: it produces dissipation and irreversibility, and possible loss
of quantum coherence.

As already mentioned, equations of the form (2.2), (2.3) have been used to 
describe various phenomena related to open quantum systems; in particular,
they have been applied to analyze the propagation and decay of neutral
meson systems.[8-13] 
Although the basic general idea behind these treatments 
is that quantum phenomena at Planck's scale produce loss of phase-coherence,
one should keep in mind that the form (2.2), (2.3) of the evolution 
equations is independent from the microscopic mechanism responsible 
for the dissipative effects.
Indeed, it is the result of very basic physical requirements 
that the complete time evolution, $\gamma_t:\rho(0)\mapsto\rho(t)$, 
needs to satisfy;
generally, the one parameter (=time) family of linear maps $\gamma_t$
should transform density matrices into density matrices and have
the properties of increasing the von Neumann entropy, 
$S=-{\rm Tr}[\rho\ln\rho]$, of obeying the semigroup composition law,
$\gamma_t[\rho(t')]=\rho(t+t')$, for $t,\, t'\geq0$, of being completely
positive.\hfill\break[1-3] In view of this, 
the equation (2.2), (2.3) can be regarded 
as phenomenological in nature; nevertheless, possible physical 
mechanisms leading these equations will be discussed in the final section.

Among the just mentioned physical requirements, complete positivity
is perhaps the less intuitive. Indeed, it has not been
enforced in previous analysis, in favor of the more obvious
simple positivity. 
Simple positivity is in fact generally enough to guarantee that
the eigenvalues of the density matrix $\rho(t)$ 
remain positive at any time; this requirement is obviously crucial
for the consistency of the formalism, in view of the interpretation of the
eigenvalues of $\rho(t)$ as probabilities.

Complete positivity is a stronger property, in the sense that
it assures the positivity of the density matrix describing the
states of a larger system, obtained by coupling in a trivial
way the neutrino system with another arbitrary 
finite-dimensional one. Although trivially satisfied by standard
quantum mechanical (unitary) time-evolutions, 
the requirement of complete positivity
seems at first a mere technical complication.
Nevertheless, it turns out to be essential in properly
treating correlated systems, like two spin-zero neutral mesons coming
from the decay of a vector-meson resonance; it assures the absence of 
unphysical effects, like the appearance of negative probabilities,
that could occur for just simply positive dynamics.[25]
For these reasons, in analyzing possible non-standard, dissipative
effects even in simpler, non correlated systems, 
the phenomenological equations (2.2) and (2.3) should always be used.%
\footnote{$^\dagger$}{We have argued before (see also the discussion
in Section 5) that the microscopic mechanism leading
to the non-standard, dissipative phenomena are likely to originate
from quantum gravity or string effects. They presumably act in the 
same way for all systems; it is therefore unjustified to adopt 
different formulations for correlated
and uncorrelated systems.}

In the case of the neutrino system, a more explicit description of
(2.2), (2.3) can be given. In the chosen basis, the effective hamiltonian
that gives rise to the standard vacuum oscillations can be written as:[22-24]
$$
H_{\rm eff}=\left(\matrix{
E+\omega&0\cr
0&E-\omega\cr}\right)\ , \eqno(2.4)
$$
where $E$ is the average neutrino energy, while $\omega=\Delta m^2/4E$ 
encodes the level splitting due to the square mass difference 
$\Delta m^2$ of the two mass eigenstates. 
In the case of oscillations in matter, 
$H_{\rm eff}$ has a more complicated expression, that takes into account
the coherent interactions of the neutrinos with the matter constituents.
For simplicity, in the following we shall limit our discussion to vacuum
oscillations: we are in fact interested in studying possible 
dissipative effects, which are quite independent from the specific 
form of the standard effective hamiltonian.

The explicit expression of the term $L[\rho]$ in (2.3) can be most 
simply given by expanding the $2\times2$ matrix $\rho$
in terms of Pauli matrices $\sigma_i$ and the identity $\sigma_0$:
$\rho=\rho_\mu\, \sigma_\mu$, $\mu=\,0$, 1, 2, 3.
In this way, the map $L[\rho]$ can be represented by
a symmetric $4\times 4$ matrix $\big[L_{\mu\nu}\big]$, 
acting on the column vector with components $(\rho_0,\rho_1,\rho_2,\rho_3)$.
It can be parametrized by the six real 
constants $a$, $b$, $c$, $\alpha$, $\beta$, and $\gamma$:[8]
$$
\big[L_{\mu\nu}\big]=-2\left(\matrix{0&0&0&0\cr
                                     0&a&b&c\cr
                                     0&b&\alpha&\beta\cr
                                     0&c&\beta&\gamma\cr}\right)
\ ,\eqno(2.5)
$$
with $a$, $\alpha$ and $\gamma$ non-negative.
These parameters are not all independent; the condition of
complete positivity of the time-evolution 
$\rho\rightarrow\rho(t)$ imposes the following inequalities:
$$
\eqalign{
&2\,R\equiv\alpha+\gamma-a\geq0\ ,\cr
&2\,S\equiv a+\gamma-\alpha\geq0\ ,\cr
&2\,T\equiv a+\alpha-\gamma\geq0\ ,\cr
&X\equiv RST-2\, bc\beta-R\beta^2-S c^2-T b^2\geq 0\ .
}\hskip -1cm
\eqalign{
&U\equiv RS-b^2\geq 0\ ,\cr
&V\equiv RT-c^2\geq 0\ ,\cr
&Z\equiv ST-\beta^2\geq 0\ ,\cr
&\phantom{\beta^2}\cr
}\eqno(2.6)
$$

Taking into account that the equation in (2.2) is trace preserving,
from the initial normalization condition ${\rm Tr}[\rho(0)]=1$, one 
immediately obtains that $\rho_0=1/2$, for all times. Then, the evolution 
equation for the remaining three components of $\rho(t)$ can
be compactly rewritten in a Schr\"odinger-like form:
$$
{\partial\over\partial t} |\rho(t)\rangle=-2\, {\cal H}\ |\rho(t)\rangle
\eqno(2.7)
$$
where the vector $|\rho(t)\rangle$ has components $(\rho_1,\rho_2,\rho_3)$,
and
$$
{\cal H}=\left(\matrix{a & b+\omega & c\cr
                       b-\omega & \alpha & \beta\cr
                       c & \beta & \gamma}\right)\ .\eqno(2.8)
$$
The formal solution of (2.7) involves the exponentiation of
the matrix $\cal H$:
$$
|\rho(t)\rangle= M(t)\ |\rho(0)\rangle\ ,\qquad
M(t)=e^{-2\,{\cal H}\,t}\ .\eqno(2.9)
$$

Let us assume that at the beginning the neutrinos were of type
``$\nu_e$''; the probability of having a transition into the type
``$\nu_\mu$'' at time $t$ is given in our formalism by:
$$
{\cal P}_{\nu_e\to\nu_\mu}(t)={\rm Tr}[\rho_{\nu_e}(t)\ \rho_{\nu_\mu}]\ ,
\eqno(2.10)
$$
where $\rho_{\nu_e}(t)$ is the solution of (2.7) with the initial condition
given by the matrix $\rho_{\nu_e}$ in (2.1). Using (2.9) and the matrices
in (2.1), one explicitly finds:
$$
{\cal P}_{\nu_e\to\nu_\mu}(t)={1\over2}\bigg\{
\cos^22\theta\, \big[1-M_{33}(t)\big] 
+\sin^2 2\theta\, \big[1-M_{11}(t)\big]
-{1\over2}\sin 4\theta\, \big[M_{13}(t)+M_{31}(t)\big]\bigg\}\ .
\eqno(2.11)
$$
When the additional piece $L[\rho]$ in (2.3) is not present, one simply
obtains:
$$ M_{11}(t)=\cos(2\,\omega t)\ ,\qquad M_{13}(t)+M_{31}(t)=\,0\ ,
\qquad M_{33}(t)=1\ ,\eqno(2.12)
$$
so that (2.11) reduces to the well known standard expression for the
oscillation probability in vacuum:
$$
{\cal P}^{(0)}_{\nu_e\to\nu_\mu}(t)=\sin^2 2\theta\ \sin^2\omega t\ .
\eqno(2.13)
$$
Therefore, any deviation from (2.12) that might be found in fitting the
expression (2.11) with data from neutrino experiments would
provide evidence for dissipative phenomena in neutrino physics.%
\footnote{$^\dagger$}{Different physical mechanisms have been proposed
in the literature to account for the observed neutrino flux deficit:
they all predict expressions for the transition probability
${\cal P}_{\nu_e\to\nu_\mu}(t)$ that differ from that in (2.11);
see the discussion at the end of Section 5.}

\vskip 2cm

{\bf 3. TRANSITION PROBABILITY: GENERAL PROPERTIES}
\medskip

Explicit expressions for the entries of the matrix $M(t)$ appearing in (2.9)
can be given by diagonalizing $\cal H$ in (2.8); this can always be done
by solving the corresponding eigenvalue equation,
$$
{\cal H}\, |v^{(k)}\rangle =\lambda^{(k)}\, |v^{(k)}\rangle\ ,
\qquad k=1,2,3\ ,\eqno(3.1)
$$
via Cardano's formula.[26] Then, using the diagonalizing matrix
$[D_{\ell k}]\equiv v^{(k)}{}_\ell$, built with the components of
the eigenvectors $|v^{(k)}\rangle$, one formally writes:
$$
M_{ij}(t)=\sum_{k=1}^3 e^{-2\lambda^{(k)}t}\ D_{ik}\, D^{-1}_{kj}\ .
\eqno(3.2)
$$

The explicit expressions of $\lambda^{(k)}$ and $[D_{\ell k}]$ 
in terms of the
dissipative parameters $a$, $b$, $c$, $\alpha$, $\beta$, $\gamma$ and
$\omega$ is however cumbersome, making the formula (3.2) unmanageable
in practice; for this reason, we shall discuss particularly interesting 
limits of the general expression (2.11) for the transition probability
${\cal P}_{\nu_e\to\nu_\mu}(t)$ in the next section.
Nevertheless, general conclusions on the behaviour of (3.2) can be
obtained by studying in more detail the eigenvalue problem in (3.1).

The three eigenvalues $\lambda^{(1)}$, $\lambda^{(2)}$, $\lambda^{(3)}$
of the matrix $\cal H$ are solutions of the cubic equation:
$$
\lambda^3+r\, \lambda^2+ s\, \lambda +w=\,0\ ,\eqno(3.3)
$$
with real coefficients,
$$
\eqalignno{
&r\equiv -(\lambda^{(1)}+\lambda^{(2)}+\lambda^{(3)})=
-(a+\alpha+\gamma)\ , &(3.4a)\cr
&\null\cr
&s\equiv \lambda^{(1)}\lambda^{(2)}+\lambda^{(1)}\lambda^{(3)}+
\lambda^{(2)}\lambda^{(3)}=a\alpha + a\gamma + \alpha\gamma
-b^2-c^2-\beta^2+\omega^2\ , &(3.4b)\cr
&\null\cr
&w\equiv -\lambda^{(1)}\lambda^{(2)}\lambda^{(3)}=
a\beta^2+\alpha c^2+\gamma(b^2-\omega^2)-a\alpha\gamma-2\,bc\beta\ .
&(3.4c)\cr}
$$
According to the sign of the associated discriminant
${\cal D}=p^3+q^2$, $p=s/3-(r/3)^2$, $q=(r/3)^3-rs/6+w/2$,
the eigenvalues are either all real (${\cal D}\leq 0$), or one
is real and the remaining two are complex conjugate (${\cal D}> 0$).
The degenerate case ${\cal D}=\,0$ occurs when two real eigenvalues
are equal; all three coincide for $p=q=\,0$.

Furthermore, the quantum dynamical semigroup generated by (2.2),
(2.3) is bounded for any $t$,[27] so that the real parts of
$\lambda^{(1)}$, $\lambda^{(2)}$, $\lambda^{(3)}$ are surely non-negative
(otherwise the entries $M_{ij}(t)$ in (3.2) would blow up for large
times).

When $\omega=\,0$, the matrix $\cal H$ is real, symmetric and 
non-negative, as guaranteed by the inequalities (2.6); therefore, 
its eigenvalues are all real and non-negative:
${\cal D}<0$ and this is possible only for $p<0$. 
The discriminant $\cal D$ starts becoming positive
only for sufficiently large $\omega$, since, as it is clear from
the definitions (3.4), the contribution of $\omega$ to $p$ is
equal to $\omega^2/3$, and thus it is positive.

Therefore, the time-behaviour of the transition probability
${\cal P}_{\nu_e\to\nu_\mu}(t)$ depends on the relative magnitude of
$\omega$ with respect to the non-standard parameters
$a$, $b$, $c$, $\alpha$, $\beta$ and $\gamma$. In particular, an oscillatory
behaviour is possible only when the dissipative parameters are small
compared to $\omega$; on the other hand, when dissipation is the
dominant phenomenon, the time-dependence in (3.2), and therefore
in (2.11), is characterized by exponential dumping terms.

This analysis allows a general discussion on the asymptotic
behaviour of ${\cal P}_{\nu_e\to\nu_\mu}(t)$ for large $t$.[27, 1] In the
generic case, ${\rm det}({\cal H})\neq0$ and all three eigenvalues
$\lambda^{(1)}$, $\lambda^{(2)}$, $\lambda^{(3)}$ are thus non-vanishing,
with positive (or zero) real part, as discussed above.
When ${\cal D}\leq0$, the eigenvalues are all real, so that
all entries of the matrix $M(t)$ in (3.2) approach zero for large $t$,
due to the exponential dumping factors.
The same is true also in presence of two complex conjugate eigenvalues, 
unless their real part is identically zero. However, this situation
never occurs when there is a non-vanishing dissipative contribution (2.5)
in the equation (2.2).
Indeed, from (3.4) one finds that the condition for having two purely
imaginary eigenvalues is given by: $w-rs=\,0$; recalling the definitions
in (2.6), it can be rewritten as:
$X +(R+S+T)[U+V+Z]+2\,(R+S+T)^3+\omega^2 (R+S+2T)=\,0$.
Since by the inequalities in (2.6) all the terms in the l.h.s. are non 
negative, they must be zero separately, which is possible only for
$a=b=c=\alpha=\beta=\gamma=\,0$.

Therefore, in presence of dissipative phenomena, the generic large $t$ 
behaviour of the transition probability in (2.11)
is independent from the mixing angle $\theta$:
$$
{\cal P}_{\nu_e\to\nu_\mu}(t)\ \asymptotic\sim\over{t\to\infty}\
{1\over2}\ .\eqno(3.5)
$$

The situation might be different however when ${\rm det}({\cal H})=\,0$
and we are in presence of zero eigenvalues. In this special case,
$\omega$ and the dissipative parameters 
$a$, $b$, $c$, $\alpha$, $\beta$ and $\gamma$
need to satisfy the additional cubic condition
$w=\,0$. Keeping $\omega$ arbitrary, the only way to satisfy this
constraint is to set $\gamma=\,0$; indeed, the inequalities (2.6)
immediately imply: $b=c=\beta=\,0$ and $a=\alpha$ and therefore
a vanishing $w$. The matrix
$\cal H$ in (2.8) takes now a very simple form, and the non-vanishing
eigenvalues are complex: $\lambda^{(1)},\lambda^{(2)}=\alpha\pm i\omega$.
Since $\alpha$ is positive, most of the entries of the evolution
matrix $M(t)$ in (3.2) are still exponentially suppressed for
large $t$; however, the presence of the zero eigenvalue now implies 
$M_{33}(t)=1$, so that the asymptotic form of (2.11) changes:
$$
{\cal P}_{\nu_e\to\nu_\mu}(t)\ \asymptotic\sim\over{t\to\infty}\
{1\over2}\sin^2 2\theta\ .
\eqno(3.6)
$$

The large-time behaviors (3.5) and (3.6), for the particular case
$\gamma=\,0$, are characteristic of the presence of the dissipative
contribution (2.5) to the evolution equation (2.2). However, in general,
it might be very difficult to distinguish these behaviours from
the one obtained in the standard case. Although in principle
${\cal P}^{(0)}_{\nu_e\to\nu_\mu}(t)$ in (2.13) has a purely oscillatory
form, in any actual observational condition, the oscillations are likely
to be averaged away, so that also in this case (3.6) holds.
Therefore, when the mixing is maximal ($\sin^2 2\theta\approx 1$), or
in the special situation in which only one dissipative parameter
is non-vanishing ($\gamma=\,0$), the asymptotic large $t$ behaviors
(3.5) and (3.6) turn out to be indistinguishable from that
of ${\cal P}^{(0)}_{\nu_e\to\nu_\mu}(t)$. In these cases, one has to study
the full time dependence of the transition probability.

\vskip 2cm

{\bf 4. TRANSITION PROBABILITY: EXPLICIT FORM}
\medskip

The general expression of the transition probability
${\cal P}_{\nu_e\to\nu_\mu}(t)$ in terms of $\omega$ and the 
dissipative parameters is very complicated and not particularly useful in
practical applications. Therefore, we shall now discuss some approximations
for which ${\cal P}_{\nu_e\to\nu_\mu}(t)$ assumes a more manageable form;
it might be of interest to compare these expressions with actual
experimental data in order to put limits on the magnitude of the
dissipative constants. Although this is clearly beyond the scope of the
present investigation, we shall nevertheless briefly comment
about the rough sensitivity that one might expect from the analysis of
present and future experiments.

As discussed before, in general ${\cal P}_{\nu_e\to\nu_\mu}(t)$ contains two
kind of contributions: oscillating terms, controlled by $\omega$,
and exponentially dumping terms, signaling dissipative effects.
The relative dominance of these two types of behaviour depends on the
magnitude of $a$, $b$, $c$, $\alpha$, $\beta$ and $\gamma$ when compared
to $\omega$. 

In our approach, the dissipative contribution (2.5)
to the evolution equation (2.2) should be regarded as phenomenological;
it is therefore hard to give an apriori estimate of the magnitude of the
non-standard parameters in $L[\rho]$. 
As mentioned in the Introduction and further discussed in the next section,
a general framework in which dissipative effects naturally emerge is
provided by the study of open quantum systems, {\it i.e.} systems
in weak interactions with large environments. In such cases
the dissipative effects can be roughly estimated to be proportional
to the typical energy scale of the system, while suppressed by inverse
powers of the characteristic energy scale of the environment.[1-3, 16, 7]

In the case of the neutrino system, on the basis of these considerations
and in line with the idea that dissipation is induced by quantum effects
at Planck's scale, one expects the values of the parameters
$a$, $b$, $c$, $\alpha$, $\beta$ and $\gamma$ in (2.5) to be very small;
for any fixed neutrino source and observational conditions,
an upper bound on the magnitude of these parameters can be 
roughly evaluated to
be of order $E^2/M_P$, with $M_P$ the Planck mass. 
The ratio of $a$, $b$, $c$, $\alpha$, $\beta$ and $\gamma$ with $\omega$
can thus be estimated to be at most of order $10^{-10}\ E^3/\Delta m^2$,
with $E$ expressed in MeV and the neutrino square mass difference
$\Delta m^2$ in ${\rm eV}^2$.
By taking for $E$ and $\Delta m^2$ values that are typical of
various neutrino sources, this ratio turns out to be about
$10^2$ for atmospheric neutrinos, of order one for solar neutrinos,
while for accelerator neutrinos it can be as small as $10^{-2}$.

When the dissipative, non-standard parameters are large or of the same
order of magnitude of $\omega$, all entries of $\cal H$ in (2.8) are
in general different from zero. In this case a useful approximation
is to assume $c$ and $\beta$ to be much smaller than the remaining
constants.%
\footnote{$^\dagger$}{Note that this choice is perfectly compatible
with the constraints of complete positivity given in (2.6)}
To lowest order, the matrix $\cal H$ becomes block diagonal and a manageable
expression for the transition probability in (2.11) can be obtained.
Explicitly, one finds:
$$
{\cal P}_{\nu_e\to\nu_\mu}(t)={1\over2}\bigg\{
\cos^2 2\theta\, \big[1-e^{-2\gamma t}\big]+
\sin^2 2\theta\, \bigg[1-e^{-At}\bigg(\cos(2\,\Omega_0t)+
{{\cal R}e B\over2\,\Omega_0}\sin(2\,\Omega_0 t)\bigg)\bigg]\bigg\}\ ,
\eqno(4.1)
$$
where
$$
A=\alpha+a\ ,\qquad B=\alpha-a+2ib\ ,\qquad
\Omega_0=\sqrt{\omega^2-|B|^2/4}\ .\eqno(4.2)
$$
The oscillating behavior in (4.1) depends on the magnitude of $\omega$
with respect to $|B|$; when $\omega<|B|/4$, 
the frequency $\Omega_0$ becomes purely
imaginary and ${\cal P}_{\nu_e\to\nu_\mu}(t)$ contains only exponential
terms. In any case, the exponential dumping terms in (4.1) dominate
for large $t$, and the limit (3.5) is recovered.

A further simplification occurs when $\gamma=\,0\,$; as already observed in
the previous section, this automatically guarantees $c=\beta=\,0$,
an further imposes $b=\,0$ and $a=\alpha$. In this case, (4.1) reduces to:
$$
{\cal P}_{\nu_e\to\nu_\mu}(t)={1\over2}\sin^2 2\theta\,
\Big[1-e^{-2\alpha t}\cos(2\,\omega t)\Big]\ .\eqno(4.3)
$$
This is the most simple form that the transition probability formula
takes in presence of dissipative effects: with respect to the standard 
expression in (2.13), (4.3) contains an exponential dumping factor in front
of the oscillating term. It can be used to derive the rough 
order of magnitude bound on the non-standard parameter $\alpha$
that can be expected from neutrino experiments.
Assuming that the dumping due to the exponential term is not exceeding
a few percent, from (4.3) one derives: $\alpha t\leq 1$.
Since the neutrinos are relativistic, the flight-time between emission and
detection is roughly the same as the distance $\ell$ between source
and detector. Then, one has: $\alpha\leq 1/\ell$, where
$1/\ell$ is approximately $10^{-22}\ {\rm GeV}$, 
$10^{-24}\ {\rm GeV}$, $10^{-27}\ {\rm GeV}$
for accelerator, atmospheric, solar neutrinos, respectively.
Although the best bound on $\alpha$ seems to be given by solar neutrinos
experiments, due to the larger $\ell$, atmospheric neutrinos data are
the most suitable for a meaningful fit of (4.3), since in this case
its time (or $\ell$) dependence can actually be probed.

Another very useful approximation of the general formula (2.11) for
the transition probability can be obtained when the non-standard parameters
$a$, $b$, $c$, $\alpha$, $\beta$ and $\gamma$ are small compared with $\omega$.
In this case, the additional dissipative term $L[\rho]$ in (2.2) can be
treated as a perturbation. Then, up to second order in the small parameters,
one explicitly gets:
$$
\eqalign{
{\cal P}_{\nu_e\to\nu_\mu}(t)={1\over2}\bigg\{&
\cos^2 2\theta\, \bigg[ 1-e^{-2\gamma t}\bigg(1+
{2\,|C|^2\over\Omega^2}\,\sin^2(\Omega t)\bigg)\bigg]\cr
+&\sin^2 2\theta\, \bigg[1-e^{-At}\bigg(\cos(2\,\Omega t) +
{{\cal R}e B\over 2\,\Omega}\sin(2\,\Omega t)-
{2({\cal I}m C)^2\over\Omega^2}\sin^2(\Omega t)\bigg)\bigg]\cr
+&\sin 4\theta\ e^{-A t}\bigg[ {{\cal R}e C\over\Omega}\sin(2\,\Omega t)
+{{\cal R}e[C(A+B-2\gamma)]\over\Omega^2}\sin^2(\Omega t)\bigg]\bigg\}\ ,}
\eqno(4.4)
$$
where $A$ and $B$ are as in (4.2), while:
$$
C=c+i\beta\ ,\qquad \Omega=\sqrt{\omega^2-|C|^2-|B|^2/4}\ .\eqno(4.5)
$$
In the previous formula, we have reconstructed the exponential
factors by consistently putting together the terms linear and 
quadratic in $t$; a similar treatment has allowed writing the oscillatory
contributions in terms of the frequency $\Omega$.%
\footnote{$^\dagger$}{This frequency is now real, since by
hypothesis $\omega^2\gg |C|^2+|B|^2/4$.}

As a further check, note that the expression (4.4) reduces to 
that in (4.2) for $|C|=\,0$, {\it i.e.} when $c=\beta=\,0$: 
it is therefore a correction to (4.2) for nonvanishing $C$.
In this respect, the validity of (4.4) goes beyond the approximation
in which it has been derived, since it can be considered as the expansion
up to second order of the general formula (2.11) for $c$ and $\beta$
small. Therefore, it can be used with confidence in fitting experimental
data from neutrino oscillation experiments.

In this respect, the data on atmospheric neutrinos are presently the best
place to look for dissipative effects. Applying techniques similar
to the ones employed {\it e.g.} in [28] and [29] to the generalized
transition probability (4.4), one should be able
to extract from the actual data useful bounds on some of the
non-standard parameters in (2.5). Nevertheless, one should note
that having in general six additional unknowns to fit will certainly make
the procedure much more difficult and complex than in the
standard case, where only the mixing angle $\theta$ and the mass
difference $\Delta m^2$ are present; only for the simplified expression
(4.3), that contains just one additional parameter besides
$\theta$ and $\Delta m^2$, one can actually expect a good fitting accuracy.

\vskip 2cm

{\bf 5. DISCUSSION}
\medskip

In the previous sections we have discussed how a phenomenological
approach based on quantum dynamical semigroup can be used to describe
dissipative dynamics for the neutrino system. As already mentioned in the 
introductory remarks, this phenomenological treatment can be supported
by physical considerations. Indeed, a general picture in which 
dissipative effects naturally emerge is provided
by systems in weak interaction with suitable environments. In the case of
elementary particle systems, these effects are likely to originate
from the dynamics of strings; however, 
an effective description of the environment, encoding some of the
``collective'' properties of the underlying fundamental theory,
is quite adequate for a more physical discussion of evolutions
of type (2.2), (2.3).[7]

To be more specific, in the case of neutrino systems, the total
hamiltonian can always be decomposed as:
$$
H_{\rm tot}=H_{\rm eff}\otimes {\bf 1} + {\bf 1}\otimes H_{\cal E} + 
g\,H'\ ,
\eqno(5.1)
$$
where $H_{\rm eff}$ is as in (2.4), while $H_{\cal E}$ describes
the internal dynamics of the environment $\cal E$. The interaction terms
between the two systems are assumed to be weak: they
are encoded in $H'$, with $g$ a small coupling constant.

Furthermore, the mechanism of neutrino production is different from
the one responsible for the dissipative effects; it is therefore natural
to assume that the neutrino state and that of 
the environment be uncorrelated
at the moment of the neutrino emission. In other words, the initial state
of the total system can be taken to be in factorized form:
$\rho_{\rm tot}=\rho\otimes\rho_{\cal E}$.

The time evolution of the neutrino state $\rho$, obtained by
tracing over the environment degrees of freedom,
\null
$$
\rho\mapsto\rho(t)=
{\rm Tr}_{\cal E}\Big[ e^{-iH_{\rm tot} t}\,
\big(\rho\otimes\rho_{\cal E}\big)\, e^{iH_{\rm tot} t}\Big]\ ,
\eqno(5.2)
$$
\null
is in general very complicated and can not be described explicitly.
Nevertheless, an evolution equation of the form (2.2), (2.3)
for $\rho(t)$ naturally emerges by taking into account the 
physical requirement that the interaction between neutrinos
and environment be weak.

There are essentially two different ways of implementing this condition:[1-3]
they correspond to the two ways of making the ratio $\tau/\tau_{\cal E}$
large, where $\tau$ is the typical variation time of $\rho(t)$, 
while $\tau_{\cal E}$
represents the typical decay time of the correlations in the environment.
Indeed, only for $\tau\gg\tau_{\cal E}$ one expects the memory effects
implicitly encoded in (5.2) to be negligible, and a local in time
evolution for $\rho(t)$ to be valid.

When $\tau_{\cal E}$ becomes small, while $\tau$ remains finite, one speaks
of ``singular coupling limit'', since the typical time-correlations of
the environment approach a $\delta$-function. In the other case, when
$\tau_{\cal E}$ remains finite, while $\tau$ becomes large, one works in the
framework of the so-called ``weak coupling limit''; in practice, this
is obtained by rescaling the time variable, $t\rightarrow t/g^2$,
and sending the coupling constant $g$ to zero (van Hove limit).

The choice between the two limits is made on the basis of physical 
considerations. In the case of unstable systems for instance, 
the weak coupling choice
is unviable, since in this case $\tau$ can be identified with the
(finite) lifetime. On the contrary, for neutrino systems both limits
are in principle allowed.%
\footnote{$^\dagger$}{Nevertheless, it should be stressed that
the condition that makes the characteristic times of the neutrino
system much larger than that of the environment, implicit in the
weak coupling limit, might not be attainable in all situations;
on the contrary, the condition on the environment time-correlations
necessary for the singular coupling limit seems more natural, in view
of its possible ``stringy'' origin.}
They give rise to different explicit expressions for the additional
contribution $L[\rho]$ in (2.3); in the case of the singular coupling
limit, one finds:
$$
L[\rho]=-\int_0^\infty dt\ {\rm Tr}_{\cal E}\bigg\{\Big[ 
e^{iH_{\cal E}t}\, H'\, e^{-iH_{\cal E}t}\ , 
\big[H',\rho\otimes\rho_{\cal E}\big]\Big]\bigg\}\ ,
\eqno(5.3)
$$
while in the weak coupling limit, one obtains:
$$
L[\rho]=-\lim_{T\to\infty}{1\over2T}\int_{-T}^T ds\ \int_0^\infty dt\ 
{\rm Tr}_{\cal E}\bigg\{ e^{iH_{\rm eff} s}\,\Big[ 
e^{iH_0 t}\, H'\, e^{-iH_0 t}\ , 
\big[H',\rho\otimes\rho_{\cal E}\big]\Big]e^{-iH_{\rm eff} s}\bigg\}\ ,
\eqno(5.4)
$$
where $H_0$ is the limit of $H_{\rm tot}$ 
when the coupling constant $g$ vanishes.

As mentioned before, the general form of the expressions for $L[\rho]$
given above does not actually depend very much on the details of the 
environment dynamics; an effective description that takes into account
its most fundamental characteristic properties is enough to allow an 
explicit evaluation of the integrals in (5.3) and (5.4).
Following the idea that the dissipative effects originate from the low
energy string dynamics at Planck's scale, one can effectively model the 
environment as a gas of D0-branes, in thermodynamic equilibrium at
Planck's temperature; these quanta obey an infinite statistics.[30-32]

Explicit computations then show that both expressions (5.3) and (5.4)
assumes precisely the form given in (2.5).%
\footnote{$^\ddagger$}{The steps followed for the evaluation of the integrals
in (5.3) and (5.4) do not much differ from the ones presented in [7];
the details are therefore omitted.}
However, while in the
case (5.3) all six parameters $a$, $b$, $c$, $\alpha$, $\beta$, $\gamma$
are in general nonvanishing, in the weak coupling limit the
average procedure in (5.4) implies $a=\alpha$ and $b=c=\beta=\gamma=\,0$,
independently from the value of $\omega$. As a consequence, when the
weak coupling limit conditions are satisfied, the dissipative
piece of the extended dynamics is controlled by a single parameter and the
transition probability ${\cal P}_{\nu_e\to\nu_\mu}(t)$ assumes the
simplified form presented in (4.3); on the other hand, the more general
behaviour (4.4) is surely the result of a singular coupling limit procedure.

Therefore, the indication of a non-vanishing value for more than one
of the parameters $a$, $b$, $c$, $\alpha$, $\beta$, $\gamma$ 
in neutrino oscillation experiments would certainly select the form
(5.3) for the dissipative piece $L[\rho]$; in turn, this would
provide some indirect informations on the structure of the environment
and thus on the effective dynamics of low energy string theory.

In closing, we would like to make a few comments on the existing literature
on the subject. Studies of possible phenomena violating quantum mechanics
in neutrino dynamics have recently appeared.[33-35] 
Based on ideas originally presented in [16], 
they discuss modifications of the standard oscillation
probability formula. However, the extended dynamics used in such
investigations is that of [16], which does not satisfy the condition
of complete positivity; as mention before, this could lead
to serious inconsistencies. We stress that to avoid these problems,
one has to adopt phenomenological descriptions based on the
equations (2.2) and (2.5).

Kinetic evolution equations similar to the one presented in Section 2
have been used to describe other, more conventional dissipative
phenomena that arise due to the scattering and absorption processes
in the core of supernovae or in the early universe.[36]
In these extreme conditions, the frequent collisions
affect the free evolutions of the neutrino species, and
the consequent decoherence effects modify the oscillation pattern.
The physical situation is now quite different from the one discussed
in the previous sections and necessarily requires a second-quantized, 
field-theoretical extension of the formalism. Further, the derivation
of the evolution equations can not rely on the weak-coupling limit
arguments discussed above; rather, it is based on the use of specific
effective interaction hamiltonians. Nevertheless, also in these cases
physical requirements like the condition of complete positivity
should in general be enforced and might turn out to be crucial
for the self-consistency of the formalism.

Dynamical equations of the form (2.2) have further been employed 
for the study of
the propagation of neutrinos in a density fluctuating media,
in particular, in the interior of the sun.[37] They give rise to
expressions for the surviving probability of the electron neutrinos
that differ from those obtained in the framework of standard matter
oscillations. Although described in terms of quantum dynamical semigroups,
these density fluctuation have their origin in the dynamics of the sun
and operate at energy scales quite different from Planck's mass.
Therefore, they can be easily isolated from the dissipative effects 
discussed in the previous sections, that, in view of their ``microscopic''
origin, are not expected to be influenced by long-range phenomena.

The recent experimental data, in particular on solar and atmospheric
neutrinos, show evidence of attenuation in the expected neutrino flux,
signaling disappearance phenomena. Although one is
led to interpret these results in terms of the standard oscillation
formula (2.13), several other physical mechanisms have been proposed
as alternative explanation for the effect, in particular: neutrino decay,
flavour changing neutral currents, violation of Lorentz invariance
or of the equivalence principle. In all these cases, the transition
probability $\cal P$ has a dependence on time (or pathlength) 
and neutrino energy that differ from the standard one. (For recent
discussions, see [28, 29].)

The dissipative effects studied here are clearly distinct and independent
from all these explanations for the neutrino flux deficit.
In particular, the dependence of $\cal P$ on the non-standard 
parameters $a$, $b$, $c$, $\alpha$, $\beta$ and $\gamma$ is distinctive of
dissipative phenomena and can not be mimicked by the other mechanisms.
This is a great advantage in the process of fitting and
comparing the experimental data,
since it makes possible the identification of the dissipative contributions
quite independently from all other effects.

\vskip 2cm

\line{\bf Note Added\hfill}
\smallskip\noindent
After the submission of our manuscript, the paper in Ref.[38] appeared;
using the atmospheric neutrino data of the Super-Kamiokande experiment
a bound on one of the dissipative parameters was obtained.

\vskip 2cm

\centerline{\bf REFERENCES}
\bigskip\medskip

\item{1.} R. Alicki and K. Lendi, {\it Quantum Dynamical Semigroups and 
Applications}, Lect. Notes Phys. {\bf 286}, (Springer-Verlag, Berlin, 1987)
\smallskip
\item{2.} V. Gorini, A. Frigerio, M. Verri, A. Kossakowski and
E.C.G. Surdarshan, Rep. Math. Phys. {\bf 13} (1978) 149 
\smallskip
\item{3.} H. Spohn, Rev. Mod. Phys. {\bf 53} (1980) 569
\smallskip
\item{4.} W.H. Louisell, {\it Quantum Statistical Properties of Radiation},
(Wiley, New York, 1973)
\smallskip
\item{5.} C.W. Gardiner, {\it Quantum Noise} (Springer, Berlin, 1992)
\smallskip
\item{6.} M.O. Scully and M.S. Zubairy, 
{\it Quantum Optics} (Cambridge University Press, Cambridge, 1997)
\smallskip
\item{7.} F. Benatti and R. Floreanini, Ann. of Phys. {\bf 273} (1999) 58
\smallskip
\item{8.} F. Benatti and R. Floreanini, Nucl. Phys. {\bf B488} (1997) 335
\smallskip
\item{9.} F. Benatti and R. Floreanini, Phys. Lett. {\bf B401} (1997) 337
\smallskip
\item{10.} F. Benatti and R. Floreanini, Nucl. Phys. {\bf B511} (1998) 550
\smallskip
\item{11.} F. Benatti and R. Floreanini, Testing complete positivity
in the neutral kaon system, in {\it CPT and Lorentz Symmetry},
V.A. Kostelecky, ed., (World Scientific, Singapore, 1999)
\smallskip
\item{12.} F. Benatti and R. Floreanini, Mod. Phys. Lett. {\bf A14}
(1999) 22
\smallskip 
\item{13.} F. Benatti and R. Floreanini, Phys. Lett. {\bf B465}
(1999) 260
\smallskip
\item{14.} F. Benatti and R. Floreanini, Phys. Lett. {\bf B451} (1999) 422
\smallskip
\item{15.} S. Hawking, Comm. Math. Phys. {\bf 87} (1983) 395; Phys. Rev. D
{\bf 37} (1988) 904; Phys. Rev. D {\bf 53} (1996) 3099;
S. Hawking and C. Hunter, Phys. Rev. D {\bf 59} (1999) 044025
\smallskip
\item{16.} J. Ellis, J.S. Hagelin, D.V. Nanopoulos and M. Srednicki,
Nucl. Phys. {\bf B241} (1984) 381; 
\smallskip
\item{17.} S. Coleman, Nucl. Phys. {\bf B307} (1988) 867
\smallskip
\item{18.} S.B. Giddings and A. Strominger, Nucl. Phys. {\bf B307} (1988) 854
\smallskip
\item{19.} M. Srednicki, Nucl. Phys. {\bf B410} (1993) 143
\smallskip
\item{20.} L.J. Garay, Phys. Rev. Lett. {\bf 80} (1998) 2508;
Phys. Rev. D {\bf 58} (1998) 124015
\smallskip
\item{21.} J. Ellis, N.E. Mavromatos and D.V. Nanopoulos, Phys. Lett.
{\bf B293} (1992) 37; Int. J. Mod. Phys. {\bf A11} (1996) 1489
\smallskip
\item{22.} C.W Kim and A. Pevsner, {\it Neutrinos in Physics and 
Astrophysics}, (Harwood Academic Press, 1993)
\smallskip
\item{23.} R.N. Mohapatra and P.B. Pal, {\it Massive Neutrinos in Physics
and Astrophysics}, 2nd ed., (World Scientific, Singapore, 1999)
\smallskip
\item{24.} S.M. Bilenky, C. Giunti and W. Grimus, Prog. Part. Nucl. Phys.
{\bf 43} (1999) 1
\smallskip
\item{25.} F. Benatti and R. Floreanini,
Mod. Phys. Lett. {\bf A12} (1997) 1465; 
Banach Center Publications, {\bf 43} (1998) 71; 
Comment on ``Searching for evolutions 
of pure states into mixed states in the two-state system $K$-$\overline{K}$'',
{\tt hep-ph/9806450}; Phys. Lett. {\bf B468} (1999) 287
\smallskip
\item{26.} For instance, see: M. Artin, {\it Algebra}, (Prentice Hall,
Englewood Cliffs (NJ), 1991)
\smallskip
\item{27.} K. Lendi, J. Phys. {\bf A 20} (1987) 13
\smallskip
\item{28.} P. Lipari and M. Lusignoli, Phys. Rev D {\bf 60} (1999) 013003
\smallskip
\item{29.} G.L. Fogli, E. Lisi, A. Marrone and G. Scioscia,
Phys. Rev. D {\bf 60} (1999) 053006
\smallskip
\item{30.} A. Strominger, Phys. Rev. Lett. {\bf 71} (1993) 3397
\smallskip
\item{31.} I.V. Volovich, D-branes, black holes and $SU(\infty)$ gauge theory,
{\tt hep-th/9608137}
\smallskip
\item{32.} D. Minic, Infinite statistics and black holes in Matrix theory,
Pennsylvania University preprint, 1997, {\tt hep-th/9712202}
\smallskip
\item{33.} Y. Liu, L. Hu and M.-L. Ge, Phys. Rev. D {\bf 56} (1997) 6648
\smallskip
\item{34.} C.-H. Chang, W.-S. Dai, X.-Q. Li, Y. Liu, F.-C. Ma and
Z. Tao, Phys. Rev. D {\bf 60} (1999) 033006
\smallskip
\item{35.} C.P. Sun and D.L. Zhou, Quantum decoherence effect and
neutrino oscillations, {\tt hep-ph/9808334}
\smallskip
\item{36.} G. Raffelt, G. Sigl and L. Stodolsky, Phys. Rev. Lett.
{\bf 70} (1993) 2363; G. Sigl and G. Raffelt, Nucl. Phys.
{\bf B406} (1993) 423; G. Raffelt, {\it Stars as Laboratories
for Fundamental Physics}, (Chicago University Press, Chicago, 1996)
\smallskip
\item{37.} C.P. Burgess and D. Michaud, Ann. of Phys. {\bf 256} (1997) 1;
P. Bamert, C.P. Burgess and D. Michaud, Nucl. Phys. {\bf B513} (1998) 319
\smallskip
\item{38.} E. Lisi, A. Marrone and D. Montanino,
Probing quantum gravity effects in atmospheric neutrino oscillations,
{\tt hep-ph/0002053}

\bye